\documentclass[11pt,a4paper]{article}
\usepackage{jcappub}
\usepackage{bm}
\usepackage{bbm}
\usepackage{amsmath}
\usepackage{etoolbox}
\usepackage{float}
\usepackage{mhchem}
\usepackage{etoolbox}

\usepackage{textcomp}
\usepackage{subfig}

\makeatletter
\gdef\@fpheader{}
\makeatother

\begin{document}

\makeatletter

\title{Constraining Below-threshold Radio Source Counts With Machine Learning}

\author[a,b]{Elisa Todarello} 
\author[b,c,d]{Andre Scaffidi}%
\author[a,b]{Marco Regis} 
\author[b]{Marco Taoso}

\affiliation[a]{Dipartimento di Fisica, Universit\`a di Torino, Via P. Giuria 1, 10125 Torino, Italy}
\affiliation[b]{Istituto Nazionale di Fisica Nucleare, Sezione di Torino, Via P. Giuria 1, 10125 Torino, Italy}
\affiliation[c]{Scuola Internazionale Superiore di Studi Avanzati (SISSA), via Bonomea 265, 34136 Trieste, Italy}
\affiliation[d]{INFN, Sezione di Trieste, via Valerio 2, 34127 Trieste, Italy}

\emailAdd{elisamaria.todarello@unito.it}. 
\emailAdd{ascaffid@sissa.it}
\emailAdd{marco.regis@unito.it}
\emailAdd{marco.taoso@to.infn.it}

\abstract{We propose a machine-learning-based technique to determine the number density of radio sources as a function of their flux density, for use in next-generation radio surveys.  The method uses a convolutional neural network trained on simulations of the radio sky to predict the number of sources in several flux bins. To train the network, we adopt a supervised approach wherein we simulate training data stemming from  a large domain of possible number count models going down to fluxes a factor of 100 below the threshold for source detection.
We test the model reconstruction capabilities as well as benchmark the expected uncertainties in the model predictions, observing good performance for fluxes down to a factor of ten below the threshold. This work demonstrates that the capabilities of simple deep learning models for radio astronomy can be useful tools for future surveys.}

\maketitle


\section{Introduction}
\label{sec:intro}

The determination of the number density of radio sources as a function of their flux density has been central in the understanding of the nature and evolution of extragalactic radio sources for more than 60 years~\cite{Condon:2012ug}. At radio frequencies, cosmological studies have been heavily based on debating source counts and the contribution of sources to the sky brightness temperature.

In the faintest regime, the connection of counts with the isotropic radio background is somewhat puzzling.
Indeed, an apparent bright high Galactic latitude diffuse radio zero level has been reported by different experiments (see \cite{Proceedings:2022ter} for a recent review). The ARCADE~2 collaboration highlighted the fact that such emission is significantly brighter than expected contributions both of Galactic and extragalactic origin~\cite{Fixsen:2009xn}. 
The isotropic component that can be isolated from radio images after subtracting foreground Galactic emission is a factor of a few larger than the total contribution obtained from extrapolating the number counts of extragalactic sources in the faint (observationally unreached) brightness regime.

It is well known that relevant statistical information about below-threshold cosmological source populations residing in our Universe can be inferred by studying fluctuations in astronomical images, in particular, from pixels that do not belong to detected sources~\cite{Scheuer,Condon:2012ug,2014MNRAS.440.2791V,2015MNRAS.447.2243V,Matthews2021}. A widely used approach is to compute the $n$-point correlation functions (with $n$ from 1 up to 3 or 4) and from them to infer the number density of sources as a function of flux and cosmological redshift.
For radio interferometric images, these computations are ``disturbed'' by the fact that the space of the physical description (i.e., the celestial sphere) does not correspond to the ``space'' where data are taken (as for any interferometer). The link is given by a Fourier transform, and different physical angular scales are mixed in the data. Radio imaging is nowadays well studied and developed to tackle this issue, nevertheless, there are a few intrinsic limitations (e.g., observationally, one cannot access all the scales needed in the transformation) that are difficult to overcome, especially when dealing with fluxes close to the root-mean-square (rms) noise of an image.

The machine-learning-based approach we devise in this work aims at a two-fold improvement with respect to traditional state-of-the-art approaches. On one side, we want to fully exploit the statistical power of the observational data around/below the detection threshold, e.g., without relying on a single estimator, such as the 1- or 2-point correlation, and, on the other side, we aim at mitigating observational issues, which, as we just mentioned, might be severe in the case of radio interferometric images.

For this purpose, we construct a convolutional neural network (CNN).
The CNN is trained using simulations of the radio sky that encompass diverse source number counts, with the goal
of making the algorithm extract the precise source number counts from a generic image.

The details about the architecture and the implementation of the CNN are reported in Sec.~\ref{sec:cnn}. In Sec.~\ref{sec:sim}, we describe both the models of source catalogs we considered to build theoretical skies and how the simulations are performed, including observational effects. This allows us to define the training, validation, and test sets. The results in Sec.~\ref{sec:res} evaluate the CNN performance and detail an empirical outline for uncertainty determination on our simulated testing sets as well as for future use on real-world maps.
A summary of the main achievements of the paper and an outlook concerning the application of the method to real data are presented in Sec.~\ref{sec:conc}.

\section{Analysis pipeline}
\label{sec:cnn}
\begin{figure}[t]
    \hspace{-3cm}
    \includegraphics[width=1.3\textwidth]{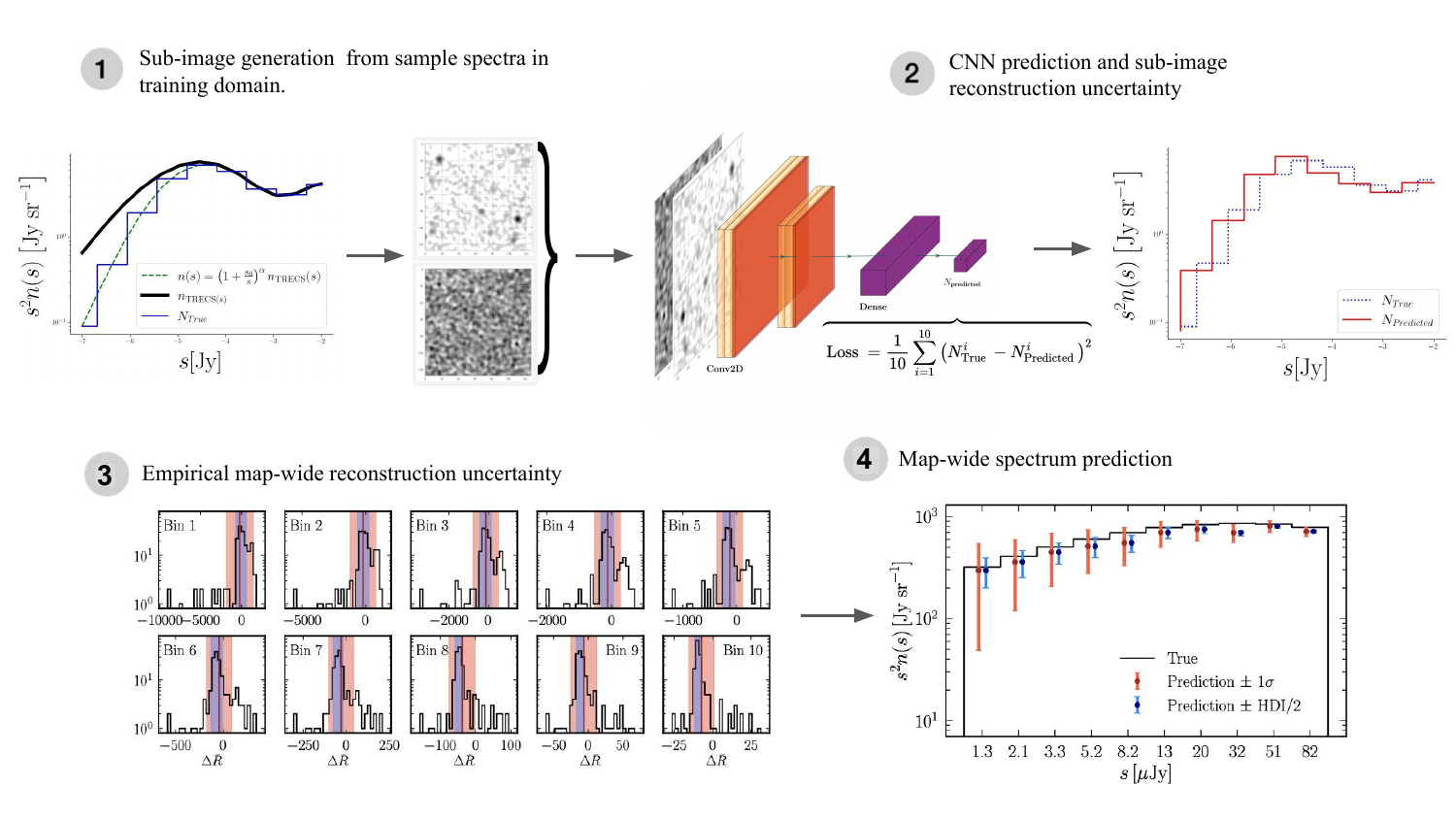}
    \caption{{A  schematic of the analysis pipeline conducted in this study. (1) Sample spectra in the training domain defined in Fig.~\ref{fig:training_set} are taken and used to generate full realised simulation maps and their residuals(as described in Sec.~\ref{sec:pipeline}) of which are divided in to semi-stochastic sub-realisations of dimentionality 128$\times$128 pixels. (2) We train a convolutional neural network (CNN) which simultaneously takes as inputs the sub-map and its corresponding residual, to output the number of sources in 10 logarithmically spaced flux bins. The neural network is trained with an MSE (L2) objective loss as shown above. The optimal architecture hyper-parameters are given in the text.  We further analyse the distribution of the CNN reconstruction capability on testing data the CNN is not trained on in order to empirically ascertain the uncertainty with which we can make predictions given real data. (3) A further 160 maps from the prior spectra domain are generated. The average map-wide reconstruction (detailed in Sec.~\ref{sec:mapwide_uncertainty}) is determined and $\pm1\sigma$ confidence and minimum width Bayesian credible interval (HDI) at 68\% containment  are calculated as estimations of uncertainty sensitivity for a real experiment. (4) Finally, pseudo-experiments are conducted and real-world analogue spectra are determined with the corresponding bin-by-bin uncertainty. }   }
    \label{fig:my_label}
    
\end{figure} 
\begin{table}[ht]
    \centering

    \begin{tabular}{||c | c | c ||}
        \hline Bin \# & Bin centers ($\mu$Jy) & Bin edges ($\mu$Jy) \\ 
        \hline  
        1 & 1.26 & [1.00, 1.58]\\  
        2 & 2.00 & [1.58, 2.51]\\  
        3 & 3.16 & [2.51, 3.98]\\  
        4 & 5.01 & [3.98, 6.31]\\  
        5 & 7.94 & [6.31, 10.00]\\ 
        6 & 12.59 & [10.00, 15.85]\\ 
        7 & 19.95 & [15.85, 25.12]\\ 
        8 & 31.62 & [25.12, 39.81]\\ 
        9 & 50.12& [39.81, 63.10]\\ 
        10 & 79.43 & [63.10, 100.00] \\ \hline
    \end{tabular}
    \caption{Flux bin centers and edges used in our supervised 10-class regression task. }
    \label{tab:bins}
\end{table}

Our goal is to construct a supervised regression task by training a CNN capable of inferring the source count at low flux densities $s$ from interferometric images and their residuals (defined in Sec.~\ref{sec:pipeline}), such as those of the Evolutionary Map of the Universe (EMU) radio survey \cite{Joseph19}. The goal is to have the network learn (and output) the source counts in 10 logarithmically spaced flux bins between $10^{-4}$ and $10^{-6}$~Jy. The bins that serve as target labels for the regression task are reported in Table~\ref{tab:bins}.  To adequately capture the features of the data and to generate outputs that are meaningful, one must carefully monitor certain aspects of neural network training. `Over-fitting' for example, is an issue that can occur when a deep neural network is sufficiently complicated that it simply learns the training data in a 1-1 fashion. This causes the network to lose all inferential capability and results in the network not being able to extrapolate to data samples that it has not seen before. To prevent this, one may stop training early whilst monitoring network performance on a separate `validation set' of data that is withheld from the training process. Another way to circumvent overfitting is to incorporate hidden layers in the network with dropout, which involves randomly dropping out (setting to zero) some fraction of a layer's neurons during training to prevent co-adaptation of neurons in a layer and to make the model more robust by forcing the network not to learn the training data exactly. Furthermore, the hyperparameters of the neural network that control the learning rate, number of hidden layers batch size, convolutional layer kernel size etc. should be optimized to achieve optimal performance. For regression tasks, this is usually done by determining the reconstruction (See. Fig~\ref{fig:result1}) for several hyperparameter set benchmarks. We detail the optimal network architecture applied to this study in Sec.~\ref{sec:arch}.

\subsection{Neural network architecture}
\label{sec:arch}
The CNN architecture is depicted in Fig.~\ref{fig:my_label} and was implemented in \texttt{Tensorflow 2}~\cite{tensorflow:2016}.
The size of the input layer is $128\times128\times 2$, where $128\times128$ is the number of pixels of each simulated image, and the factor $2$ accounts for the fact the interferometric radio images are simultaneously parsed into the network with their corresponding residuals. The training+testing set contains $\mathcal{O}(10^5)$ objects, see Sec.~\ref{sec:sim} for more details on the data generation. The input layer is followed by three hidden layers: two 2D max-pooling (orange in Fig.~\ref{fig:my_label}) and LeakyReLU (yellow) sequences with 0.25 dropout which are subsequently flattened into a dense layer with a ten-dimensional output with linear activation, corresponding to each flux bin of the regression task. The training was optimized over 33 epochs with a batch size of 100, learning rate = 0.001, and LeakyRelU negative slope coefficient of 0.2. We adopt a mean squared error (MSE) loss function and use train/validation split of 97/3\%. We monitor training by implementing an early stopping criterion to ensure no over-fitting occurs. Results of the model testing are described in Sec.~\ref{sec:res}. 


\section{Data sets}
\label{sec:sim}
We perform the simulations of the observed radio sky using the ASKAPsoft tool~\cite{Guzman}. The reason for this choice is two-fold. On one hand, even though we aim at devising a generic approach, suitable for any radio interferometric survey, we need to focus on a given telescope, in order to accurately and realistically introduce observational effects in our images. The ASKAP telescope is one of the current world-leading radio telescopes available and is accompanied by a well-developed simulation pipeline. The second reason stems from the fact that the largest (in terms of sky coverage) radio survey, ongoing at the time of writing, is provided by the EMU observations, conducted with the ASKAP telescope.
Therefore, despite the developed method will have wider applicability, we test it by simulating images similar to the ones being observed by EMU at 940 MHz, which will offer an immediate future application.

\subsection{Simulation pipeline}\label{sec:pipeline}
Our simulation pipeline consists of four steps.
Our first task is to create a suitable catalog of sources in the sky with known source counts. As a starting point, we take the Tiered Radio Extragalactic Continuum Simulation (T-RECS) simulated ``medium" catalog of extra-galactic sources \cite{Bonaldi}, scaled to a frequency of 940~MHz. This catalog spans 25~deg$^2$ and contains about 30 million sources with a flux lower limit of $10^{-8}$~Jy. We truncate the catalog to only include sources with flux densities ranging from $10^{-6}$ to $10^{-2}$~Jy. This will be approximately the range of pixel values for our input images. Larger values are possible because a pixel will in general contain multiple sources. The lower limit allows us to have catalogs with manageable file sizes, including only relevant sources, while thanks to the upper limit we avoid the presence of individual very bright sources that would require a specific cleaning procedure to mitigate side lobes, something doable for a single image obtained from observations, but not for a high number of simulated images.
{ \footnote{When dealing with observational data of a survey like EMU, one can check whether the subtraction of individual bright sources might affect the results, by comparing ``quiet'' regions with regions where such subtraction has been performed.}}

\begin{figure}[ht]
\centering
\includegraphics[width=0.7\textwidth]{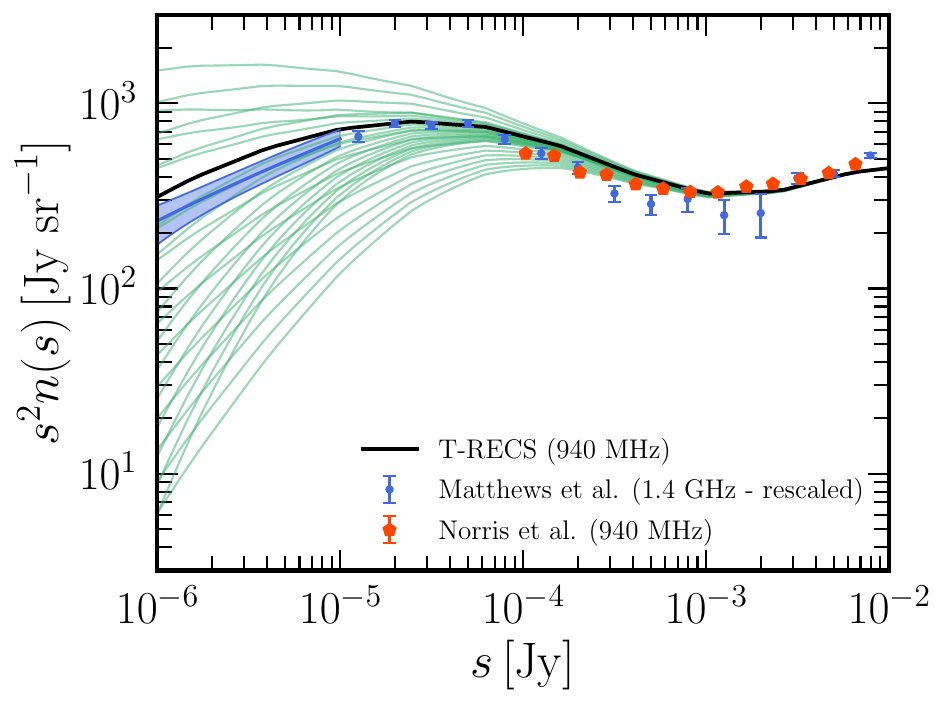}
\caption{Differential number counts used to create the training sets (green lines) as a function of the flux density. 
The T-RECS spectrum at 940 MHz is shown in black.
With red pentagons, we report the measurement of source counts of~\cite{Norris:2021yqm}, derived from observations with the same properties as those simulated here.
For reference, and in order to understand where physical models should lie, we show deeper measurements from~\cite{Matthews2021}  (blue points, derived at 1.4 GHz, rescaled here at 940 MHz), as well as the region obtained with the $\mathcal{P}(D)$ method on the same data~\cite{Matthews2021} (blue band). 
}                                   
\label{fig:training_set}
\end{figure}

The differential number count of sources extracted from the T-RECS catalog $n(s)=dN(s)/ds$ at 940~MHz is shown in black in Fig.~\ref{fig:training_set}. 
We create new catalogs for the training set with a variety of $n(s)$ by modifying the T-RECS one, as shown in Fig.~\ref{fig:training_set}. We choose the following functional form with two free parameters $\alpha$ and $s_0$
\begin{equation}
n(s) = \left( 1 + \frac{s_0}{s}\right)^\alpha n_{\rm TRECS}(s) \enspace.
\label{ns}
\end{equation}
For given $\alpha$ and $s_0$ we determine the total number of sources to be included in the simulation. Then we arbitrarily select a sample from the T-RECS catalog with the desired number of sources. We take the size of the sources from the T-RECS catalog and it turns out they are all essentially point-like for the angular resolution we are considering.
To each source, we assign a new flux at random, such that the differential source counts over the whole image reproduces Eq.~\eqref{ns} with the desired values of $\alpha$ and $s_0$.
The number of sources in any given catalog ranges from $\sim~4\times 10^4$ to  $\sim~1\times 10^7$, which means that a good sampling of the flux distribution is guaranteed.

The second step of the simulation pipeline involves turning such catalogs into a sky model, i.e., into a theoretical image with sources projected on the sky.
This is done with the tool \textit{cmodel} of the ASKAPsoft package~\cite{Guzman}, choosing to center the field at typical RA and DEC coordinates of ASKAP observations, namely RA=21h and DEC=-55\textdegree.

Next, the observation is simulated, the output being the visibilities with telescope noise included. 
We perform the simulation through the tool \textit{csimulator} by setting 6 hours of observation in continuum-wide mode with 1 channel of 288 MHz centered around 940~MHz. The antenna locations, feed locations, and other relevant definitions are taken as for the default ASKAP pipeline~\cite{Matpriv}.

In the last step, the visibilities are converted to a celestial image, the so-called dirty image.
Subsequently, deconvolution with the point spread function is performed with the CLEAN algorithm. The residuals, i.e. the difference between the CLEANed and dirty images, are also stored for use in the analysis. For all these tasks we employ the tool \textit{cimager}.
We apply a Wiener filter with robustness equal to zero, as in~\cite{Norris:2021yqm}, and a 
$30^{\prime\prime}\times30^{\prime\prime}$ Gaussian taper for preconditioning our images,  which is then approximately the size of our beam. For dealing with non-coplanarity, we apply a w-projection with 129 planes.
We run the CLEAN algorithm with scales [0,3,10] pixels and a loop gain of 0.3. We perform 3 major cycles, each consisting of a maximum of 15000 minor cycles. A major cycle is exited when a 15\% or 0.3~mJy CLEAN limit is reached.

The rms noise level of our images, determined as the rms of an image with no sources run through the pipeline described above, is 23~$\mu$Jy/beam, in good agreement with expectations~\cite{Norris:2021yqm}.
\begin{figure}
    \centering
    \includegraphics[width=0.8\textwidth]{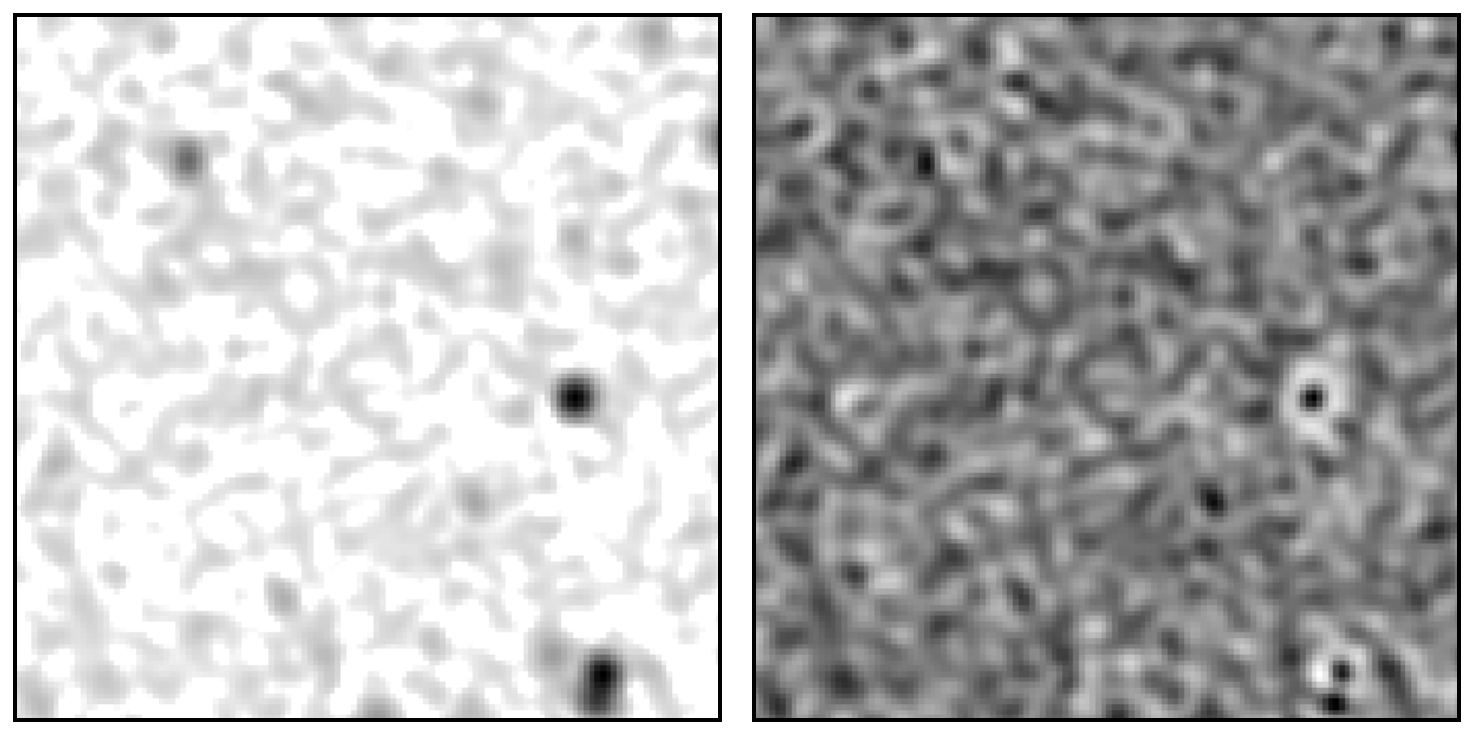} 
    \caption{Example realizations of the data used to train the CNN. \textbf{Left:} $128\times128$ sub-image, processed with normalization and stretching, 
    as described in the text in Sec.~\ref{sec:train/test}.  \textbf{Right:} Residual image obtained applying the CLEAN algorithm, as introduced in the text.
   }
    \label{fig:subimages}
\end{figure}

\subsection{Train and test sets}
\label{sec:train/test}

Our training and validation sets span 28 pairs of $\alpha$ and $s_0$, see Eq.~\ref{ns}, leading to the models depicted as green lines in Fig.~\ref{fig:training_set}. For each pair, we generate 2 source catalogs, using the Monte Carlo sampling described above. For each of the four cases with the lowest source counts, we generate 2 additional catalogs, in order to improve statistics. Furthermore, we create one catalog with $s_0=0$, corresponding to the T-RECS flux spectrum, for a total of 65 catalogs. The training set generated from the catalogs consists of 65 of 25~deg$^2$ images, each made of 2560$^2$ pixels, for a pixel size of about $7^{\prime\prime}\times 7^{\prime\prime}$. 

In Fig.~\ref{fig:training_set}, the aforementioned models are compared with three different determinations of the differential source counts from existing observations. 
Red pentagons show the counts obtained by the EMU Pilot survey at 940~MHz from~\cite{Norris:2021yqm}. As already mentioned above, the EMU observations have been producing images with properties similar to the ones simulated in this work. We see from Fig.~\ref{fig:training_set}, that the extraction of the source count with ``traditional techniques" (i.e., ``counting" detected sources) can be performed down to $\sim 10^{-4}$ Jy (i.e., about 5 times the rms noise). 
The method proposed here aims at significantly extending this range towards lower fluxes.

The blue points in Fig.~\ref{fig:training_set} show the measurement from~\cite{Matthews2021}, taken at 1.4 GHz and derived with observations deeper than the EMU ones. We scaled the measurement to 940~MHz adopting a frequency dependence $s\sim\nu^{-0.7}$.
They are plotted in order to guide our eyes toward the physical models that are more plausible. On the other hand, the blue points shouldn't be directly compared with our findings, since, as said, we are not simulating the corresponding observational setup, but the one of the EMU survey.
The same argument applies to the blue band, which shows the determination of the differential source counts obtained from the $\mathcal{P}(D)$ technique, using the same observations~\cite{Matthews2021}.
The so-called $\mathcal{P}(D)$ method~\cite{Scheuer} allows to constrain the source counts from the statistics of pixel intensities in confusion-limited radio maps, reaching fluxes down to a factor of a few below the noise level, see e.g.~\cite{Condon:2012ug,2014MNRAS.440.2791V,2015MNRAS.447.2243V,Matthews2021}. In this sense, the machine learning technique proposed in this work can be seen as alternative or complementary to a $\mathcal{P}(D)$ analysis. 
Although our results shouldn't be directly compared with the blue band, we use it to identify the ``most likely" region where the real source count is expected to lie, which we call the region of interest (ROI), and use it to define some benchmark scenarios.
Unfortunately, a similar $\mathcal{P}(D)$ band from EMU data is not available at the time of writing.

As shown in Fig.~\ref{fig:training_set}, our training models, span a large range of source counts, even beyond those allowed by the aforementioned measurements.
Large data variety is important for adequate training of the CNN. It serves to improve the ability of the network to generalize to features it has not seen before. In other words, it prevents over-fitting, which occurs when the neural network simply learns the training data~\cite{zhou2011analyses}. 

We found that significant improvement, of the order of $\sim20\%$, in reconstruction accuracy arose when additional information from the image residuals was fed to the neural network. Thus, for each realization, we consider two images, the CLEANed image and residuals. Implementing this approach was inspired by Ref.~\cite{2014MNRAS.440.2791V,2015MNRAS.447.2243V}, where information on the residuals was supplied to enhance the $\mathcal{P}(D)$ reach.

\begin{figure}[ht]
\centering
\includegraphics[width=0.7\textwidth]{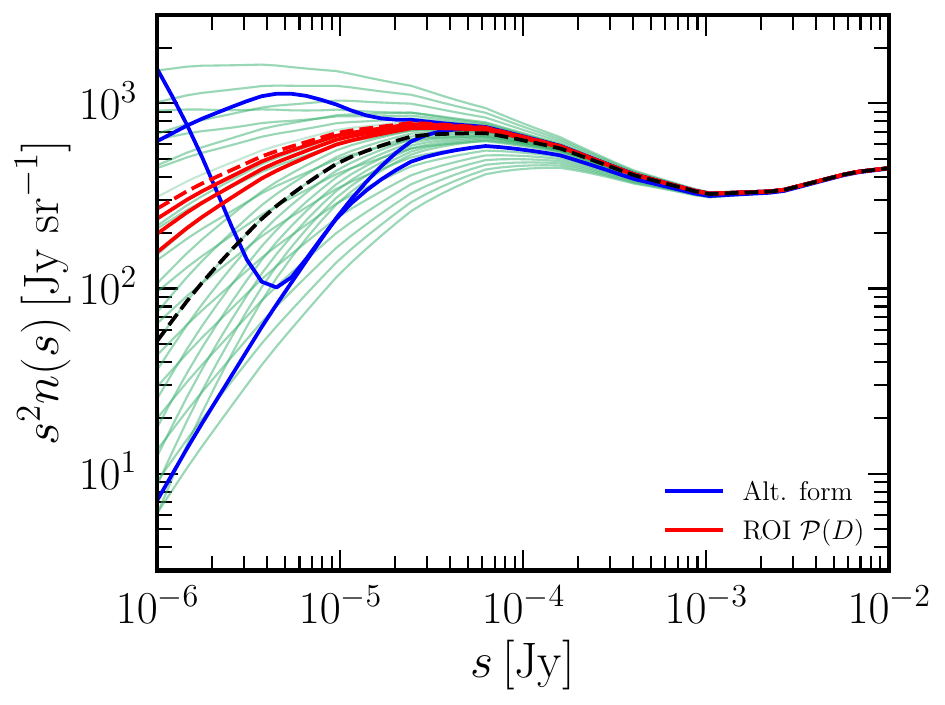}
\caption{Models of the testing set. The thick blue lines represent the test set used to ascertain the network's extrapolation ability when $n(s)$ is not of the form of Eq.~\eqref{ns}. The thick red lines (both solid and dashed) are the flux spectra falling within the $\mathcal{P}(D)$ region shown in Figure~\ref{fig:training_set}.
The thick dashed lines (one red, one black) correspond to the images  
used to recreate a real analysis pipeline with the results displayed in Figure~\ref{fig:recon}.}  
\label{fig:different}
\end{figure}

CNN's are more efficiently trained with small-size images. One must take care to optimize the available information present in the image, whilst simultaneously allowing for network performance. Therefore, we split each image into 400 sub-images, for a total of 26000 sub-images. Example realizations of these sub-images are displayed in Fig.~\ref{fig:subimages}. We show a CLEANed sub-image (left) and its corresponding residuals (right). The sub-images underwent minimal pre-processing, namely, we first normalized the images by the maximum
pixel value of the whole training set and then applied a square root stretch of all pixel values $p_i \rightarrow \sqrt{p_i}$ (where $p_i$ corresponds to the value of the flux in pixel $i$, where any negative pixel values were set to zero\footnote{All image pre-processing and manipulation was performed using the \texttt{astropy} module \cite{astropy:2022}.}).  This transformation reduces the dynamic range of pixel values in the image and was observed to maximize the regression power of the CNN\footnote{This transformation is by no means unique or optimal. However a logarithmic transformation, for example, was observed to give inferior results. }. 
Residuals were also normalized the images by the maximum
pixel value over the whole training set.
For testing the model performance in Sec.~\ref{sec:res},  we additionally generate 57 maps with spectral parameters in the same range as the training set. Moreover, we generate  100 maps sampling four indicative number count distributions that fall within the ROI around the $\mathcal{P}(D)$ shown in Fig.~\ref{fig:training_set}. Finally, we also generate three images whose flux spectra do not have the functional shape in Eq.~\ref{ns}, see Fig.~\ref{fig:different} and  Section~\ref{sec:generalize}. The total testing set then comprises 160 maps, and thus 64000 sub-images.

\subsection{Labels}
\label{sec:labels}
\begin{figure}
\centering
\includegraphics[width=0.7\textwidth]{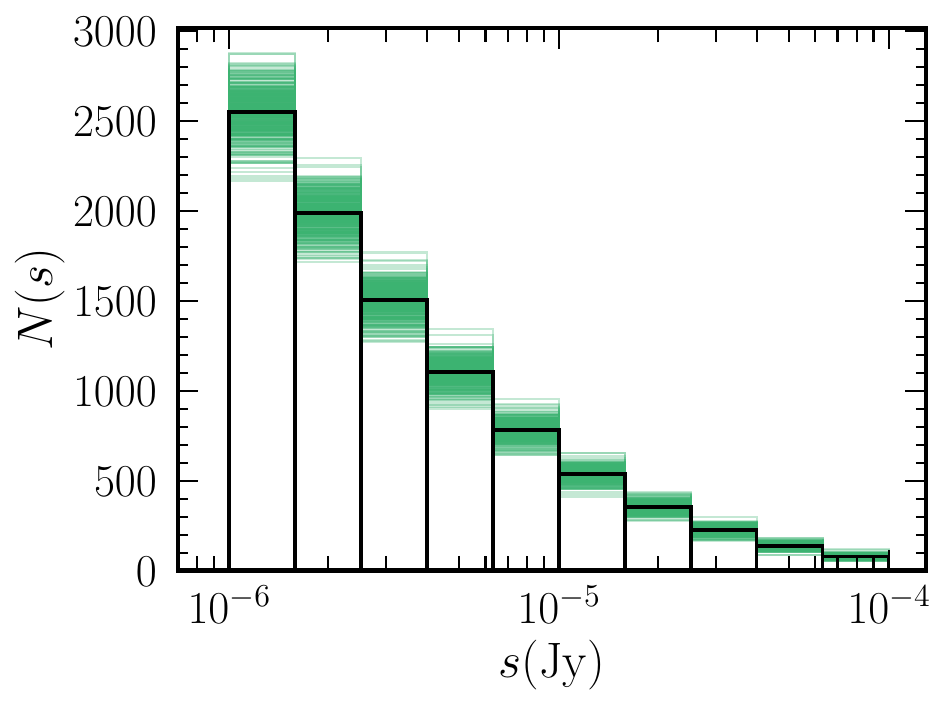}
\caption{Variability of labels among sub-images belonging to the same 25~deg$^2$ image. The green lines represent the 400 sub-images, while the thick black line is their average. The model used here is the T-RECS one, reported with a black line in Fig.~\ref{fig:training_set}.}
\label{fig:var}
\end{figure}

The CNN is trained on the sub-images discussed in the previous section. The labels for the supervised regression task are the number of sources in the 10 flux bins for each sub-image. The process of splitting the full map into 400 sub-images causes each sub-image to contain a sample of the total number of map-wide sources. The labels are then stochastically distributed around some mean (in an approximately Gaussian way), with this mean corresponding to the \textit{true} source count numbers for an entire map, as shown in Fig.~\ref{fig:var}. This distinction is important to make, since the model prediction uncertainties to be discussed in Sec. \ref{sec:res} necessarily need to take simulation uncertainties into account in a meaningful way.

Note that results are shown either in terms of $n(s)$ or $N(s)$, the two being related by: $n(s)=N(s)/\Delta s/\Delta \Omega$, where $\Delta s$ is the flux bin, see Table~\ref{tab:bins}, and $\Delta \Omega$ is the angular size of the sub-image.

\section{Results}
\label{sec:res}
\begin{figure}
\centering
\includegraphics[width=\textwidth]{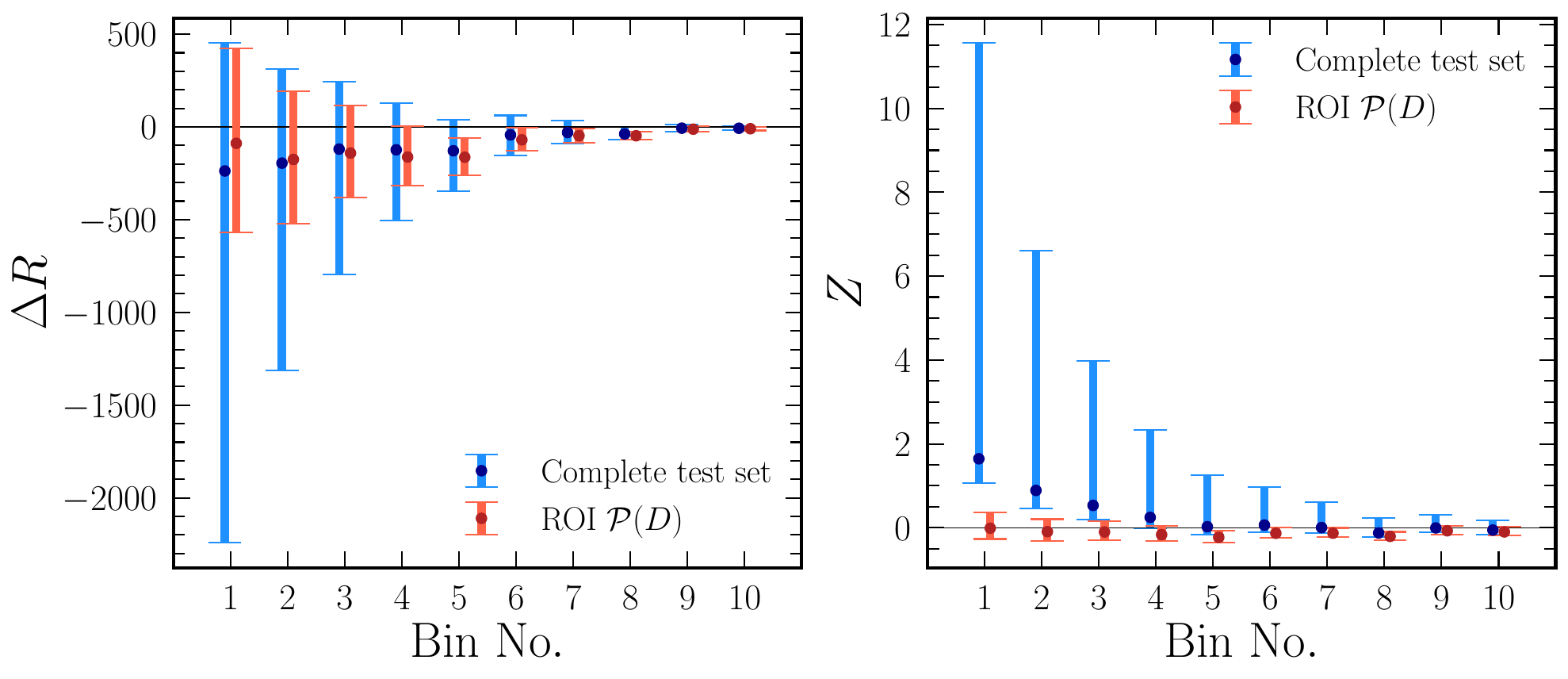}
\caption{\textbf{Left:}  Test set sub-image reconstruction residuals $\Delta R$ defined in Eq.~\ref{eq:rel_recon0}. The points correspond to the mean of the distribution with 400 $N_{set}$ entries and the error bars are computed as the standard deviation of the cases overshooting the mean (upper bar) and undershooting the mean (lower bar). The complete testing set, corresponding to $\alpha$ and $s_0$ values in the full training domain, are displayed in blue. The testing sub-set corresponding to spectra within the ROI are displayed in red.   
\textbf{Right}: The same as the left panel but for the relative reconstruction residual for each sub-image $Z$, defined in Eq.~\ref{eq:rel_recon1}. The large separation between the mean $Z$ and 0 for the first bins originates from the systematic over-reconstruction observed for models with very low counts.}
\label{fig:result1}
\end{figure}
\begin{figure}
\centering
\includegraphics[width=0.8\textwidth]{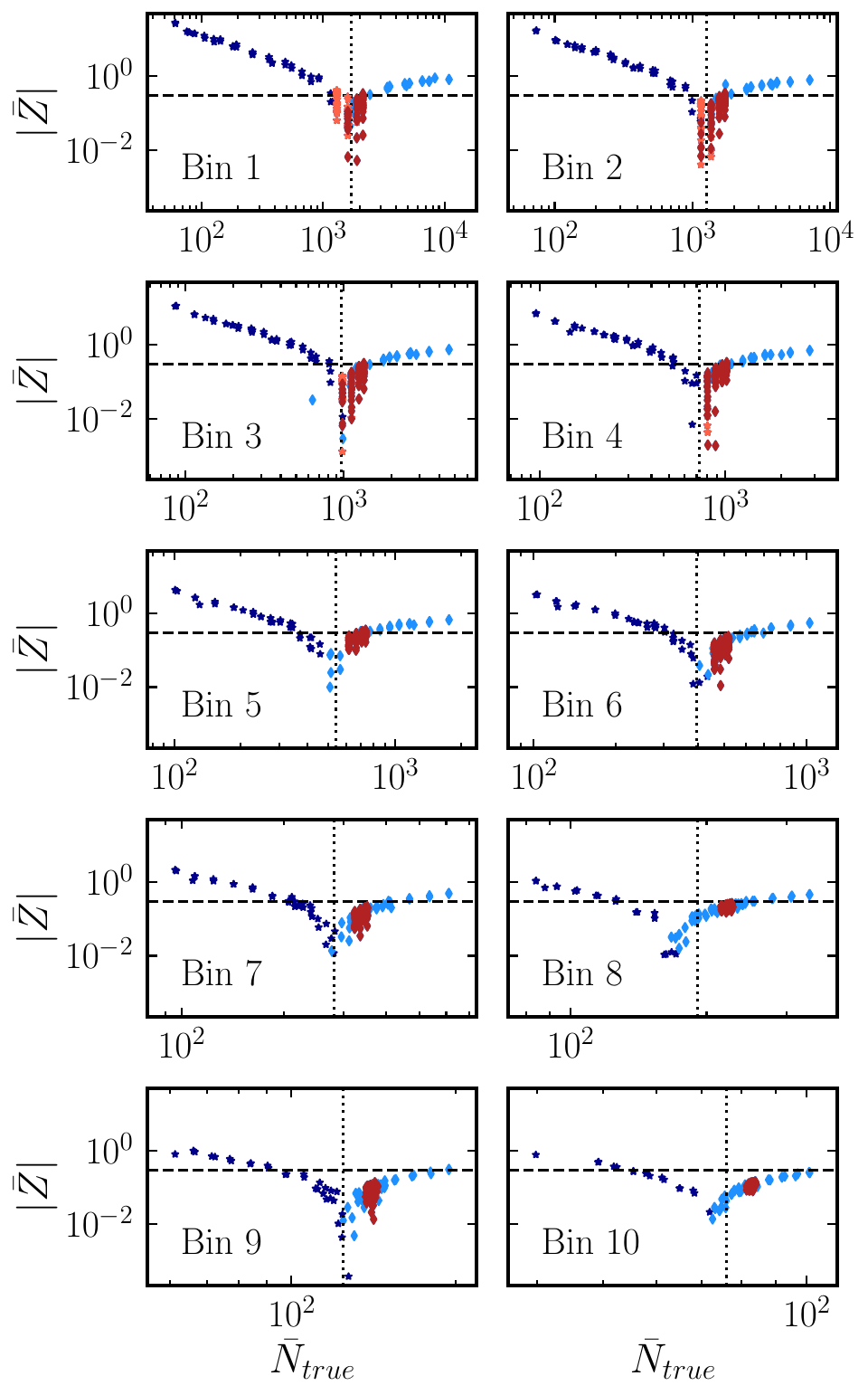}
\caption{Absolute value of the average (map-wide) relative reconstruction residuals $\bar Z$ (Eq.~\ref{eq:rel_recon}) in each target bin as a function of $\bar{N}_{true}$. 
In blue, we show $|\bar{Z}|$ for maps from the entire domain of models that were used for training, while in red we show $|\bar{Z}|$ for the 100 maps that correspond to the region surrounding the extrapolated region of $\mathcal{P}(D)$ \cite{Matthews2021}. 
Dark blue and light red stars indicate $\bar{Z} > 0$, while light blue and dark red diamonds indicate $\bar{Z}<0$.
For illustration purposes, the horizontal dashed line marks $|\bar{Z}| = 0.3$, i.e., a 30\% deviation from the true value. The vertical dotted line marks the average $\bar{N}_{true}$ over the images of the training set for each flux bin.}
\label{fig:result2}
\end{figure}

To evaluate the adequacy of the proposed approach, in view of a future application of the method to real data, we thoroughly test the model's performance both on the whole test set, as well as on the subset of test images corresponding to the ROI alone.
The prescription we adopt aims to systematically determine an estimate of the model's accuracy, namely how well it predicts the number of sources in each flux bin, as well as the uncertainty on this prediction. In this section, we first display measures of the model's reconstruction performance after being fed testing sub-image sets, followed by a detailed measure of the uncertainty on the bin-by-bin predictions for an entire map which will serve as  a more realistic test-bed for the model's performance on real data.

\subsection{Reconstruction accuracy}
A preliminary check of model performance is the reconstruction residual in each flux bin. It is shown on the left panel of Fig.~\ref{fig:result1}. Here the vertical axis shows the reconstruction in the sub-images of our test sets
\begin{equation}
\Delta R = N^{ijk}_{predicted} - N^{ijk}_{true} \enspace,
\label{eq:rel_recon0}
\end{equation}
where $i$ labels the sub-images ($i=1,..,400)$, $j$ the flux bin ($j=1,..,10)$ and $k$ the image in our test set  ($k=1,..,N_{set}$). The estimator we plot in Fig.~\ref{fig:result1} is the mean over all the images and sub-images in the test set, and the error bars are the standard deviations. We distinguish testing sets by color. The blue points correspond to the ensemble of $\alpha$ and $s_0$ values covering the whole training domain, whilst the red points correspond to the subset of test spectra that fall in our ROI, i.e., around the $\mathcal{P}(D)$ region of Fig.~\ref{fig:training_set}, shown with red lines in Fig.~\ref{fig:different}. We computed upper error bars from the standard deviation of cases overshooting the mean value and lower error bars for undershooting cases. We see that for the brightest bins the error bars are rather symmetric, which reflects the fact that the underlying distributions are fairly Gaussian. In the case of the faintest flux bins, the distributions have more pronounced tails, which lead to asymmetric errors. In these absolute terms, i.e., in terms of the direct output of the CNN, we observe a systematic increase in reconstruction accuracy in higher flux bins.
On the other hand, it is more meaningful to show the relative reconstruction accuracy, reported on the right panel of Fig.~\ref{fig:result1}, in terms of 
\begin{equation}
Z=\frac{N^{ijk}_{predicted} - N^{ijk}_{true}}{N^{ijk}_{true}} \enspace,
\label{eq:rel_recon1}
\end{equation}
showing again the average over all the images and sub-images in the test set as the prediction estimator and computing errors as the standard deviations.
Here it is evident that, when considering the entire test set and the faintest flux bins, the upper error is more pronounced than the lower error. This arises because the CNN prediction for the lowest curves in Fig.~\ref{fig:different} is significantly overshooting the true value, inducing a significant upper tail in the distribution.
In the first flux bins an average overestimation is observed in the right panel of Fig.~\ref{fig:result1}, driven by models with low counts. On the other hand, the mean value of $\Delta R$ in the left panel of Fig.~\ref{fig:result1} is dominated by models with large counts, for which the CNN tends to underestimate the true values.

In a real experiment, one is of course not interested in the predicted spectrum of each sub-image, but rather of the full aggregated 25 deg$^2$ map made by 400 sub-images. Therefore, in order to ascertain the bin-by-bin reconstruction accuracy across the simulated testing maps, we build the relative reconstruction residual in the flux bin $j$ averaged over the sub-images in a given image $k$:
\begin{equation}
\bar Z = \frac{1}{400}\sum_{i=1}^{400} \frac{N^{ijk}_{predicted} - N^{ijk}_{true}}{N^{ijk}_{true}} \enspace,
\label{eq:rel_recon}
\end{equation} 
and show it in Fig.~\ref{fig:result2} as a function of the average number of sources of each sub-image inside an image, i.e, for given $j$ and $k$,  $\bar N_{true}=\sum_{i=1}^{400}N^{ijk}_{true}/400$.

From this plot, we can verify that the worst performance is for images with few sources and for low fluxes. 
The trend for the model to have less precise reconstruction on the images that have lower statistics is the manifestation of an issue called `data scarcity' in deep learning. In this situation, the neural network's performance can suffer due to a lack of representative examples. We have done our best to circumvent this issue by providing ample training data for the model in the form of 400 sub-images per map. This indeed cures the issue for high fluxes. 
For the first flux bins, the impact of noise is severe and we reach the limit of applicability of our technique, i.e., for low fluxes and a low number of sources, it becomes too difficult for the CNN to disentangle between noise and source fluctuations, and there is a systematic over-prediction due to misinterpretation of noise fluctuations (recall that negative pixels are set to zero in the CLEANed images). However, this happens only for extreme models, corresponding to the lowest curves in Fig.~\ref{fig:training_set}.  For models in the ROI around the extrapolated $\mathcal{P}(D)$~\cite{Matthews2021}, that are shown with red markers in Fig.~\ref{fig:result2}, this issue dissipates across all bins, as the number of sources in these distributions was adequately large and representative\footnote{In general, various techniques such as data augmentation, transfer learning, and domain adaptation can be used to improve the neural network's ability to generalize to new and unseen data. We leave the use of these methods to future work. For example, domain adaptation is potentially useful in this area, where one can make use of the raw source images as the `source domain' in such an approach~\cite{farahani2020brief}.}.

Conversely, for large number of sources, the CNN tends to under-predict the counts. This becomes relevant when the number of sources is larger than $10^3$ and is due to confusion. Indeed, each sub-map of our simulations is composed of $\sim 10^3$ beams. To tackle this issue one could increase the resolution of the simulated images, but, again, this is not an issue for the physically plausible models.


Another important test we identified is to show that the neural network had not simply learned the functional form of the spectra presented in the training set (or ones very similar), which we will address later in Section~\ref{sec:generalize}.

\begin{figure}[ht]
\centering
\includegraphics[width=\textwidth]{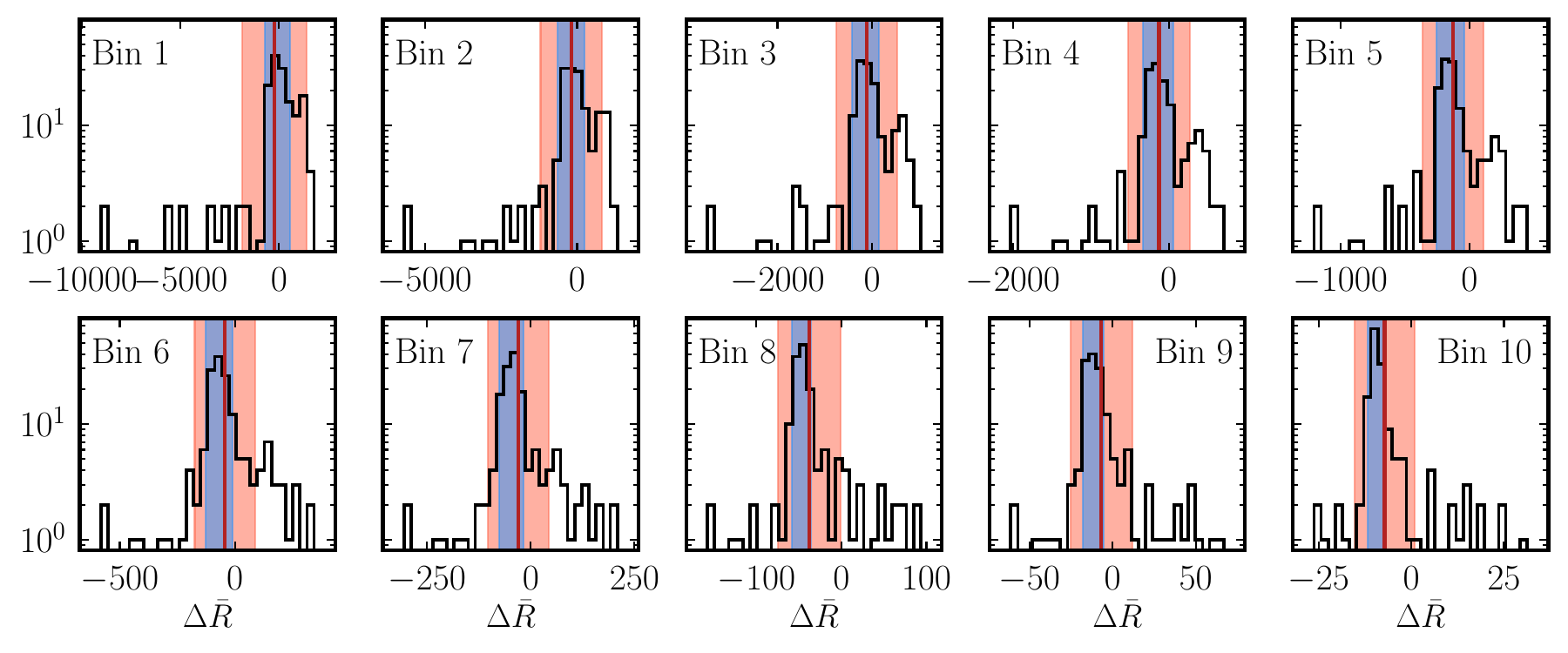}
\caption{Distributions of the average test map reconstruction for each map in the full test set (entire spectral domain). The blue bands indicate the minimum width Bayesian credible interval containment at 68\% and the light red regions
are $\pm 1$ standard deviation estimators for the samples in the distribution.  For our study, we adopt the conservative uncertainties shown by the red bands as the uncertainty in the neural network's prediction for the number of sources in a given flux bin for a given map. For illustration the red vertical line denotes the average $\Delta\bar{R}$ in each bin.}
\label{fig:mc_err}
\end{figure}


\begin{figure}[ht]
\centering
\subfloat{%
  \includegraphics[width=0.7\textwidth]{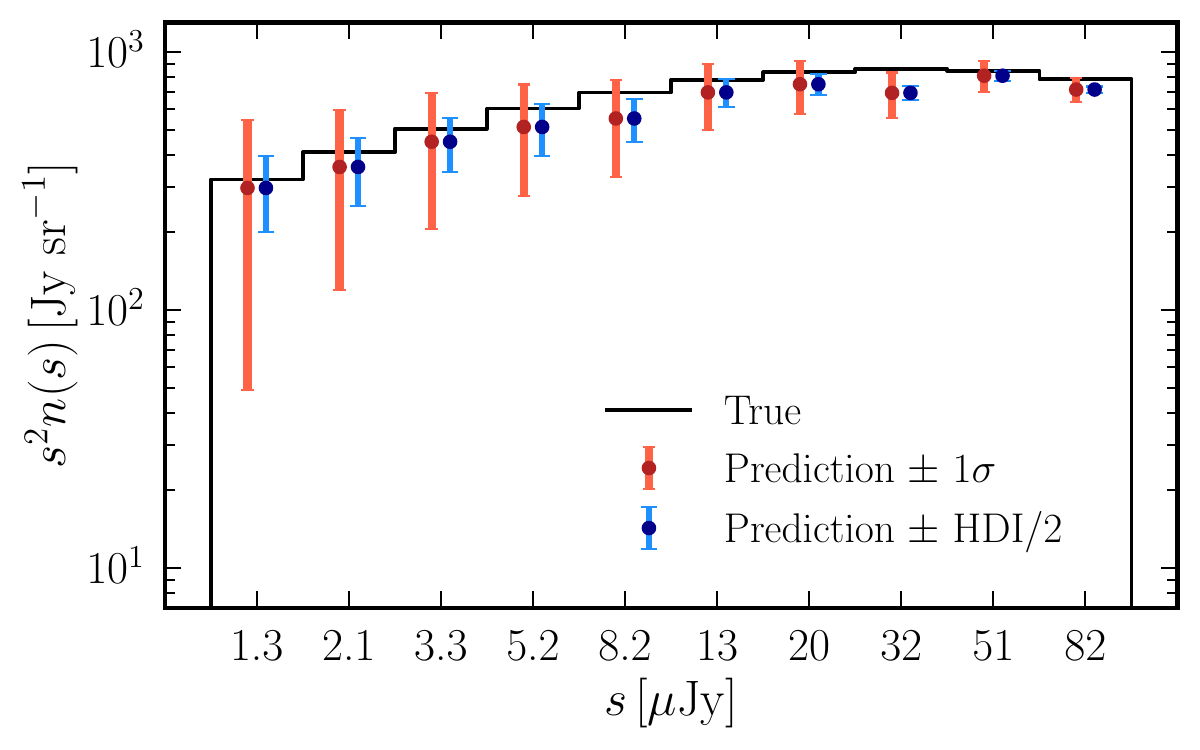}%
}

\subfloat{%
  \includegraphics[width=0.7\textwidth]{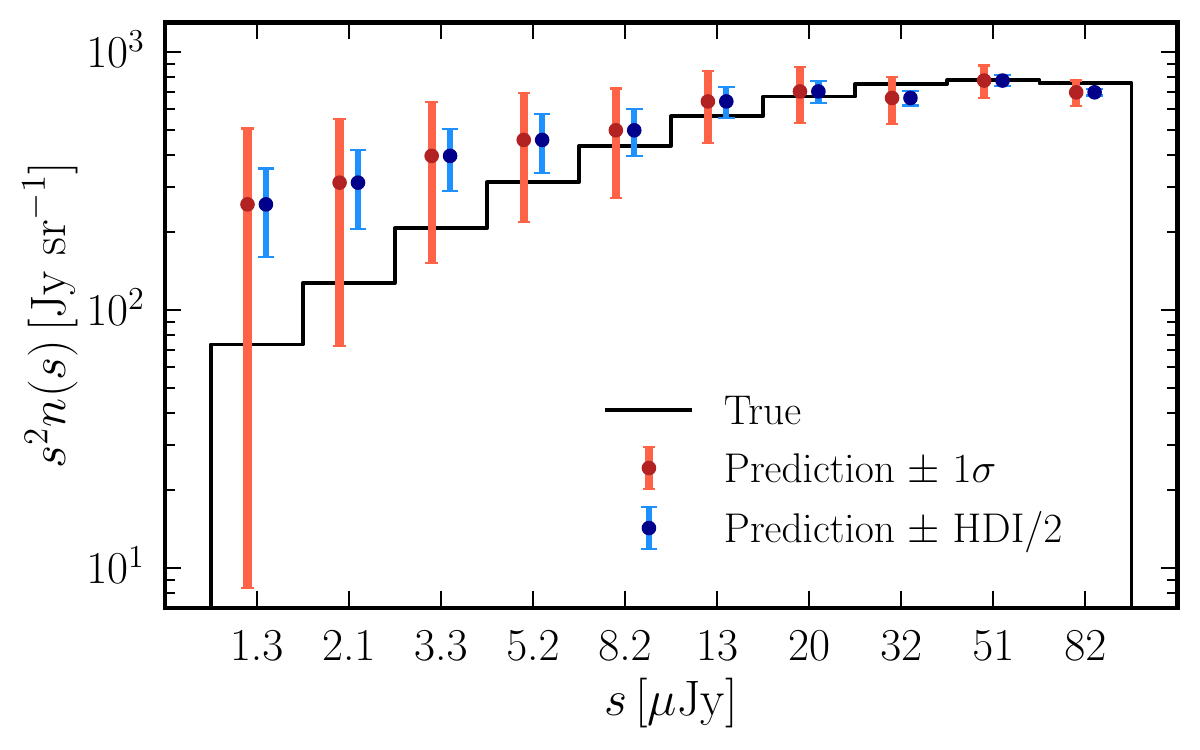}%
}

\caption{Simulated realizations of a real experiment. Shown are the predicted (dots) map-wide flux spectra with associated true labels (solid black line) corresponding to two images in the test set. The systematic map-wide label uncertainty is negligible, but accounted for by the displayed error bars. \textbf{Top:} Predicted flux spectrum for a map in the $\mathcal{P}(D)$ ROI. \textbf{Bottom}: Predicted spectrum for a map outside the ROI. The corresponding flux spectra are shown in Figure~\ref{fig:different}. The red and blue error bars are the $\pm 1\sigma$ and $\pm \mathrm{HDI}/2$ prediction uncertainties, respectively. They are derived from the red and blue contours of the distributions in Fig.~\ref{fig:mc_err}. 
}
\label{fig:recon}
\end{figure}


\subsection{Uncertainty estimation}
\label{sec:mapwide_uncertainty}
In order to more quantitatively assess the level of the uncertainties that one would have when using our network to analyze real data, we empirically build the probability distribution function of the average reconstruction error $\Delta \bar R$, i.e., of the reconstruction error of Eq.~\ref{eq:rel_recon0} averaged over the entire map, $\Delta \bar R=\sum_{i=1}^{400}(N^{ijk}_{predicted}-N^{ijk}_{true})/400$, for a given image $k$ and flux bin $j$. 
For the whole test set comprising 160 maps (64000 sub-images), covering the same range of models as the training set, see Fig.~\ref{fig:different}, we calculate the  distribution of the expected reconstruction error for each bin, by building a histogram including the $\Delta \bar R$ values of each map, in order to objectively, in the frequentist sense, determine the variance of the predictions. We assume that the representative uncertainty in our map-wide prediction is given asymptotically (in the limit of infinite representative testing data) by the uncertainty in $\Delta \bar{R}$. This ensures a conservative estimate of the network map-wide performance once all sub-images have been aggregated, and also allows to account for the systematic sampling uncertainty that arises from the uncertainty on the labels of each sub-image in the training set;  Recall, the labels in each sub-image are stochastic due to the nature of the image generation (see Sec.~\ref{sec:sim}), and so the mean reconstruction per map is also systematically affected by this source of error. The results are shown in Fig.~\ref{fig:mc_err} for each bin. The distributions are highly non-Gaussian. We therefore display two uncertainty measures, given by a) the minimum width Bayesian interval (blue bands) or Highest Density Interval (HDI), defined as the narrowest credible interval containing 68\% probability that can be constructed taking the histogram as the posterior distribution, and b) the standard deviation estimator over all maps in the test set (light red bands). The  minimum width Bayesian interval turns out to be significantly more constraining, while the standard deviation estimator is more conservative.


We then consider two arbitrary models, one belonging to the ROI and one outside, shown with bold dashed lines in Fig.~\ref{fig:different}. In Fig.~\ref{fig:recon}, we report their reconstructed counts, showing the Bayesian and frequentist-derived uncertainties as error bars.
To see the impact of the label systematic sampling uncertainty mentioned above on the overall error, we compute it as the standard deviation of the mean true count value: $\sigma_L=\sqrt{\sum_{i=1}^{400}(N^{ijk}_{true}-\bar N^{jk}_{true})^2}/400$. It turns out to be negligible if compared to the error bars in Fig.~\ref{fig:recon}.

Fig.~\ref{fig:recon} demonstrates that the CNN is performing well, at least in the ROI, and that the estimate of the uncertainty is reliable/conservative when considering the standard deviation.

\begin{figure}
    \centering
    \includegraphics[width=0.8\textwidth]{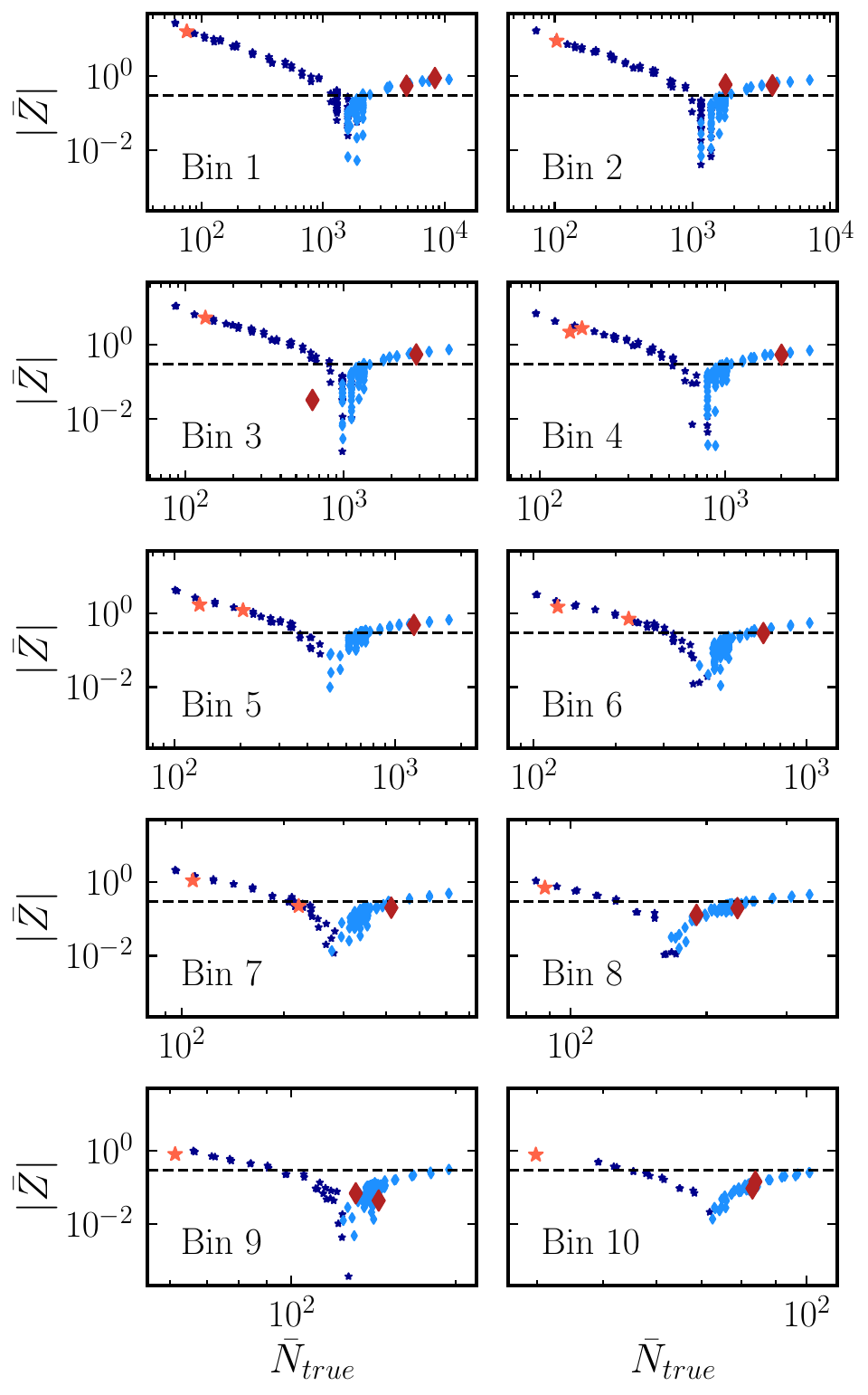}
    \caption{The same as Fig.~\ref{fig:result2}, but the red markers now correspond to the test spectra shown with bold blue lines in Fig.~\ref{fig:different}, with highly non-trivial functional forms. 
    The reconstruction performance at low source count follows the same trends as the rest of the testing set. As in Fig.~~\ref{fig:result2}, dark blue and light red stars indicate $\bar{Z} > 0$, while light blue and dark red diamonds indicate $\bar{Z}<0$.
    }
    \label{fig:3_spec}
\end{figure}
\subsection{Verification of model generalization}\label{sec:generalize}
To test the model's extrapolation capability and coverage, we included three maps with spectra that deviate significantly from any of the functional forms of the training set. These models are shown with bold blue lines in Fig.~\ref{fig:different}. We verified that the CNN can reliably reconstruct the number counts for those cases, that are drastically different from what it has seen before. We show the result in Fig.~\ref{fig:3_spec}, where the new models are reported with red makers, while blue markers are the same as in Fig.~\ref{fig:result2}, reported for comparison. We observe that their reconstruction accuracy aligns with the one of the models with $\alpha$ and $s_0$ as in the training set.

\section{Conclusions and outlook}
\label{sec:conc}

In this work, we have investigated a machine-learning-based approach to extract the differential number counts of sources from interferometric radio maps. 
In order to probe the feasibility of the method we have trained and tested a CNN  on simulated radio images. 
More specifically, we have employed the ASKAPsoft tool~\cite{Guzman} to simulate EMU observations at 940 MHz performed by the ASKAP telescope. 
The synthetic images have an rms noise of 23~$\mu$Jy/beam and have been generated for a variety of differential number count distributions, shown in Fig.~\ref{fig:training_set}. Each simulated image span 25~deg$^2$ and has been divided into 400 sub-images, representing different statistical realizations of the underlying source distribution.
A total of 26000 sub-images has been used for training and validation and a sample of 64000 sub-images for testing.
The algorithm has been trained in order to reconstruct the source counts in 10 flux bins between $10^{-6}$ and $10^{-4}$~Jy.
The measurement of the number counts from sources detected in the EMU Pilot survey~\cite{Norris:2021yqm}, shown in Fig.~\ref{fig:training_set}, reaches fluxes down to $\sim10^{-4}$~Jy, which corresponds roughly to 5 times the rms noise of EMU observations~\cite{Norris:2021yqm}.  




The goal of our analysis is to investigate whether a machine-learning approach can be used to improve the determination of the source counts with respect this `traditional' method, extending the reach at lower fluxes.

Our results can be understood from Figs.~\ref{fig:result2}, \ref{fig:mc_err} and \ref{fig:recon}. We have found that for source count distributions favored by a $\mathcal{P}(D)$ analysis at 1.4 GHz~\cite{Matthews2021} (see the blue band in Fig.~\ref{fig:training_set} and red stars in Fig.~\ref{fig:result2}), the CNN algorithm can reconstruct the source counts with an accuracy better than $\sim$30\% in the whole flux range probed by our analysis.
For much lower (higher) counts, the accuracy significantly degrades because of the impact of noise (confusion), especially in the low flux bins.
From the distributions in Fig.~\ref{fig:mc_err}, we have derived the count uncertainty in each flux bin to be assigned to a generic measurement.
In units of the counts of physically motivated models, such as the T-RECS one (black curve in Fig.~\ref{fig:training_set}) or the ones favored by the $\mathcal{P}(D)$ (red curves in Fig.~\ref{fig:different}), such uncertainty is $\lesssim 30\%$ down to $10^{-5}$ Jy (i.e., down to fluxes that are one order of magnitude below the threshold in~\cite{Norris:2021yqm}) and better than a factor of 2 down to $10^{-6}$ Jy, see upper panel in Fig.~\ref{fig:recon}.
These results show that the proposed machine-learning method provides a good reconstruction of the source count distribution in the unresolved regime, with performance comparable (or possibly better) than the $\mathcal{P}(D)$ technique~\cite{Condon:2012ug}.
On the other hand, as shown in Fig.~\ref{fig:result2}, the CNN algorithm tends to predict a level of source counts close to the one favored by the $\mathcal{P}(D)$ analysis,  leading to a systematic under-prediction (over-prediction) of the counts for models with a large (small) number of sources. We plan to investigate further to this behavior in future work, in order to improve the accuracy of our machine-learning method. Within the current analysis, such a trend is accounted for by our estimate of the uncertainty of the predictions, and therefore of the  performance of the CNN algorithm.

Finally, a thorough  comparison and investigation of the complementarity with the $\mathcal{P}(D)$ analysis will be the next methodological step for assessing the usefulness of the machine-learning method.
On the observational side, the EMU Main Survey has started, and images are currently being collected. They will be the arena where to test the proposed CNN approach on real data in the immediate future.



\section*{Acknowledgements}
We are indebted to Paulus Lahur and Matthew Whiting for their help in running ASKAPsoft.
We would like to thank S. Ammazzalorso and L. Thomas for their contribution in the initial phase of the project, and D. Parkinson and T. Vernstrom for useful discussions. AS is supported by the programme "DS4ASTRO: Data Science methods for Multi-Messenger Astrophysics \& Multi-Survey Cosmology”, in the framework of the PRO3 Programma Congiunto (DM n. 289/2021) of the Italian Ministry for University and Research. MT acknowledges the research grant ``The Dark Universe: A Synergic Multimessenger Approach No. 2017X7X85'' funded by MIUR. ET, AS, MR and MT acknowledge the project ``Theoretical Astroparticle Physics (TAsP)'' funded by Istituto Nazionale di Fisica Nucleare (INFN). MR and ET acknowledge support from `Departments of Excellence 2018-2022' grant awarded by the Italian Ministry of Education, University and Research (\textsc{miur}) L.\ 232/2016 and Research grant `From Darklight to Dark Matter: understanding the galaxy/matter  connection to measure the Universe' No.\ 20179P3PKJ funded by \textsc{miur}. ET thanks the Galileo Galilei Institute for Theoretical Physics for its hospitality during the completion of this work.

\bibliographystyle{JHEP_improved}
\bibliography{refs.bib}

\end{document}